\documentclass{jpsj-suppl}
\usepackage{txfonts} 
\title{Tunka-Rex: the Cost-Effective Radio Extension of the Tunka Air-Shower Observatory}
\author{
F.G.~\textsc{Schr\"oder}$^{1}$, 
P.~\textsc{Bezyazeekov}$^{2}$, 
N.M.~\textsc{Budnev}$^{2}$, 
O.A.~\textsc{Gress}$^{2}$, 
A.~\textsc{Haungs}$^{1}$, 
R.~\textsc{Hiller}$^{1}$, 
T.~\textsc{Huege}$^{1}$, 
Y.~\textsc{Kazarina}$^{1,2}$, 
M.~\textsc{Kleifges}$^{3}$, 
E.N.~\textsc{Konstantinov}$^{2}$, 
E.E.~\textsc{Korosteleva}$^{4}$, 
D.~\textsc{Kostunin}$^{1}$, 
O.~\textsc{Kr\"omer}$^{3}$, 
L.A.~\textsc{Kuzmichev}$^{4}$, 
R.R.~\textsc{Mirgazov}$^{2}$, 
L.~\textsc{Pankov}$^{2}$, 
V.V.~\textsc{Prosin}$^{4}$, 
G.I.~\textsc{Rubtsov}$^{5}$, 
V.~\textsc{Savinov}$^{2}$, 
R.~\textsc{Wischnewski}$^{6}$, 
A.~\textsc{Zagorodnikov}$^{2}$ 
}

\inst{
$^{1}$Institut f\"ur Kernphysik, Karlsruhe Institute of Technology (KIT), Germany\\
$^{2}$Institute of Applied Physics ISU, Irkutsk, Russia\\
$^{3}$Institut f\"ur Prozessdatenverarbeitung und Elektronik, Karlsruhe Institute of Technology (KIT), Germany\\
$^{4}$Skobeltsyn Institute of Nuclear Physics MSU, Moscow, Russia\\
$^{5}$Institute for Nuclear Research of the Russian Academy of Sciences, Moscow, Russia\\
$^{6}$DESY, Zeuthen, Germany\\
}

\email{frank.schroeder@kit.edu}

\abst{Tunka-Rex is the radio extension of the Tunka cosmic-ray observatory in Siberia close to Lake Baikal. Since October 2012 Tunka-Rex measures the radio signal of air-showers in coincidence with the non-imaging air-Cherenkov array Tunka-133. Furthermore, this year additional antennas will go into operation triggered by the new scintillator array Tunka-Grande measuring the secondary electrons and muons of air showers. Tunka-Rex is a demonstrator for how economic an antenna array can be without losing significant performance: we have decided for simple and robust SALLA antennas, and we share the existing DAQ running in slave mode with the PMT detectors and the scintillators, respectively. This means that Tunka-Rex is triggered externally, and does not need its own infrastructure and DAQ for hybrid measurements. By this, the performance and the added value of the supplementary radio measurements can be studied, in particular, the precision for the reconstructed energy and the shower maximum in the energy range of approximately $10^{17}-10^{18}\,$eV. Here we show first results on the energy reconstruction indicating that radio measurements can compete with air-Cherenkov measurements in precision. Moreover, we discuss future plans for Tunka-Rex.}

\kword{UHECR 2014, Tunka-Rex, radio, cosmic ray air shower}

\begin{document}
\maketitle

\section{Introduction}
For more than 10 years, digital radio measurements of air showers have been under investigation as alternative detection technique for ultra-high-energy cosmic rays \cite{FalckeNature2005, ArdouinBelletoileCharrier2005}. The radio technique has to compete in performance and price with established techniques: First, the detection of secondary air shower particles at ground, which enables measurements around the clock, but suffers from statistical fluctuations and imperfections in the hadronic interaction models used for interpretation of the measurements. Second, the detection techniques relying on light emitted by the air-shower, namely measurements of air-Cherenkov light used until energies around $10^{18}\,$eV, and air-fluorescence light used for energies above $10^{17}\,$eV. These methods using light detection feature a precise calorimetric measurement of the shower energy, and a relatively precise measurement of the shower maximum, but are limited to dark nights with clear weather, i.e., duty cycles in the order of $10\,\%$ (after quality cuts).

Radio measurements are feasible around the clock, as particle measurements, and share the advantage of a calorimetric measurement of the shower energy \cite{LOPES_XmaxLDF2014}. Moreover, radio measurements are sensitive to the longitudinal shower development \cite{LOPES_PRD2012}. Recent results of the dense LOFAR array indicate that radio measurements can compete in the precision for the reconstruction of the shower maximum \cite{LOFAR_Xmax2014}. Thus, radio measurements combine the principle advantages of the established detection techniques: high precision and high duty cycle. However, exploiting these advantages requires an array which is significantly denser than corresponding arrays of particle detectors. Moreover, self-triggering on the radio signal is difficult, which makes radio measurements more suitable for hybrid arrays featuring additional detectors providing the trigger. Thus, an important question is for which type of measurements and science goals, radio arrays or radio extensions to particle detector arrays yield the most cost-effective option.

Tunka-Rex is a cost-effective radio extension to the existing Tunka-133 air-Cherenkov array \cite{Tunka133_NIM2014}. With additional costs of approximately $500\,\$$ per antenna station, Tunka-Rex adds radio measurements to high-energy events measured by Tunka-133. The primary goal of Tunka-Rex is a cross-calibration of the radio and the air-Cherenkov signal. By direct comparison, we can test the precision of the radio reconstruction for the energy and for the shower maximum. Afterwards, the additional radio information could be used to improve the total reconstruction accuracy of the hybrid events, e.g., by exploiting the paradigm of shower universality. In near future, additional Tunka-Rex antennas already deployed will be triggered by the new over- and underground scintillator array Tunka-Grande, which will increase the duty cycle and the statistics around $10^{18}\,$ eV by an order of magnitude.

\begin{figure}
\centering
\includegraphics[width=0.70\textwidth]{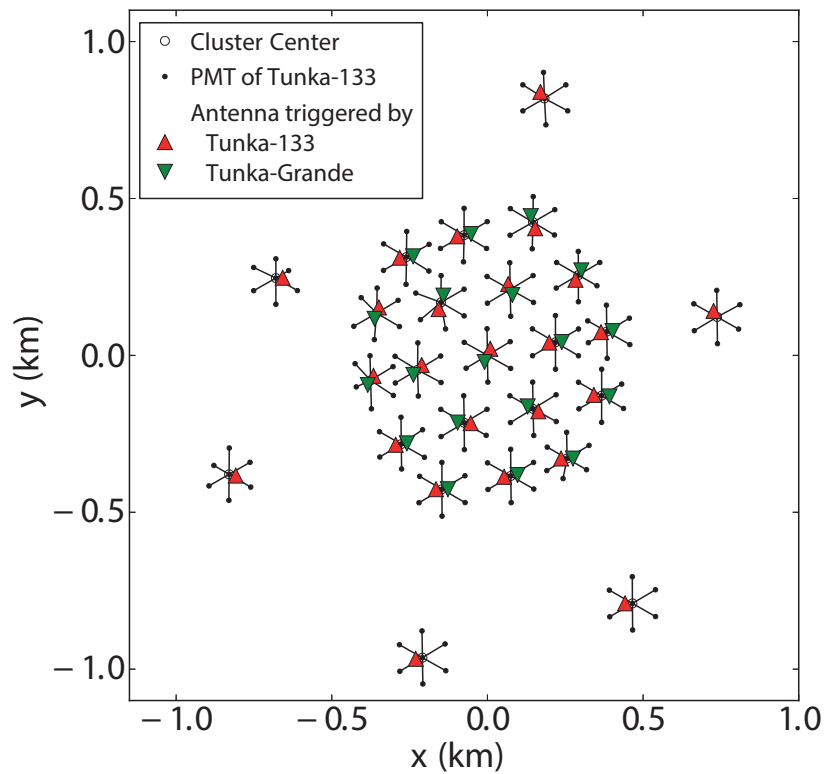}
\caption{Map of Tunka-Rex. Operation started in 2012 with antennas triggered by the air-Cherenkov array of photo-multiplier (PMT) detectors. The additional antennas triggered by the scintillator array Tunka-Grande are already deployed, end will go into operation this year.} \label{fig_map}
\end{figure}

\section{Experimental setup}
Tunka-Rex makes extensive use of the existing infrastructure at the Tunka observatory. In particular Tunka-Rex shares the existing data-acquisition system (DAQ) of the air-Cherenkov array Tunka-133. Tunka-133 meanwhile consists of 25 clusters covering an area of approximately $3\,$km\textsuperscript{2}. Each cluster features its own local DAQ, and contains 7 non-imaging photo-multiplier detectors and one Tunka-Rex antenna station. Thus, whenever Tunka-133 triggers an event, automatically the radio signal is recorded, too. While the cost of a stand-alone radio array would be dominated by infrastructure and data-acquisition, this design makes Tunka-Rex a cost-effective extension for radio-hybrid measurements.

Tunka-Rex started operation in October 2012 with 17 antennas, and was extended in autumn 2013 to 25 antennas (see figure \ref{fig_map}). Meanwhile Tunka-Rex has measured for approximately 800 hours. This corresponds to an effective duty-cycle of approximately $5\,\%$, because Tunka-Rex is only running when Tunka-133 is operating. To increase the duty-cycle, one additional antenna has been installed at each of the inner 19 clusters, which will be triggered by the new particle-detector array Tunka-Grande. Tunka-Grande is an array built from former KASCADE-Grande scintillators. It has one station per cluster consisting of surface detectors, and underground detectors for muons. Step-by-step, it went already partially in operation, and will provide the trigger and shared DAQ required for an increased duty-cycle of Tunka-Rex.

Each Tunka-Rex antenna station consists of two orthogonally aligned antennas (i.e. two channels), which enables a reconstruction of the radio polarization. The antennas are $20\,$m distant from any other detector and the local DAQ at each cluster center, to avoid direct radio interferences. Coaxial cables feed the signal into a filter-amplifier at the local DAQ where the signals of both antenna channels are digitized simultaneously with the photo-multiplier signals. The filter restricts the signal to a bandwidth of $30-80$\,MHz, a frequency range widely used for air-shower detection, since it provides a relatively high signal-to-noise ratio: at higher frequencies the radio signal is generally weaker, at lower frequencies galactic noise dominates.

Since Tunka-Rex aims at demonstrating that radio measurements can be done economically, also the analog detector components of Tunka-Rex have been optimized for cost: As antenna type we have chosen the SALLA, a simple and robust antenna with integrated low-noise preamplifier. Compared to other antenna types it has a lower gain \cite{AERA_AntennaPaper2012}, but as advantage over other antenna types, the SALLA is almost independent of ground conditions. This means that Tunka-Rex has a slightly higher threshold by using the SALLA, but also lower systematic uncertainties, since ground conditions (in particular humidity) vary drastically in the Tunka valley with the seasons. Also for cost-optimization, the filter-amplifiers are in-house productions based on a proprietary design developed for the Auger Engineering Radio Array (AERA) \cite{AERA_Reference}.

\begin{figure}[t]
\centering
\includegraphics[width=0.99\textwidth]{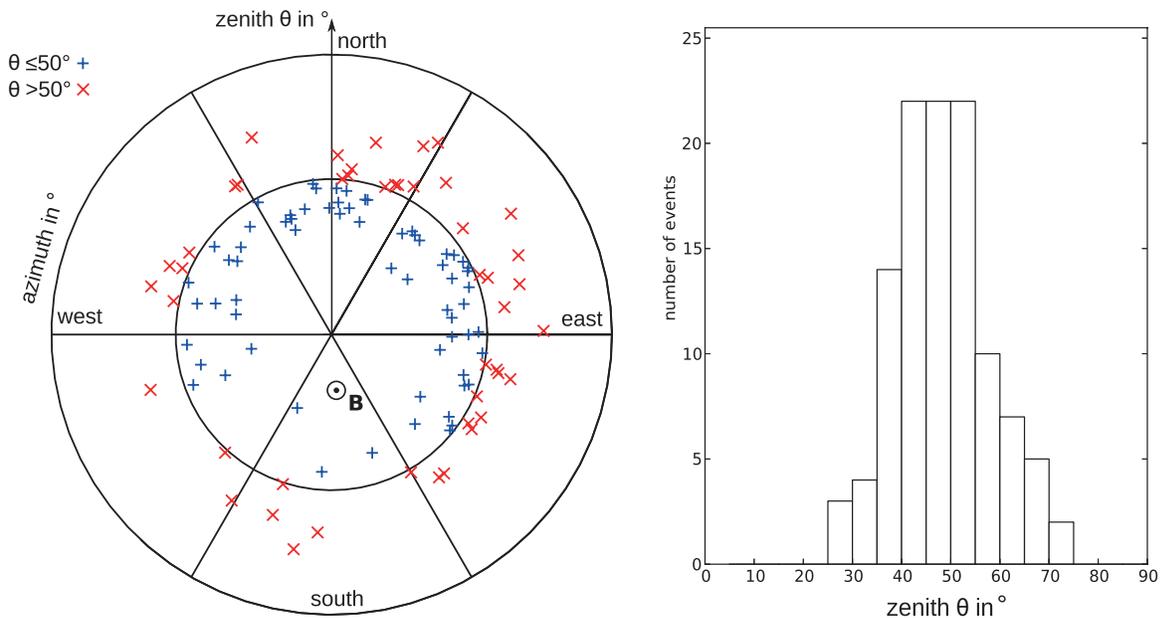}
\caption{Sky map and distribution of the shower inclination of Tunka-Rex events. For events with zenith angles $\theta \le 50^\circ$, Tunka-133 provides a reconstruction of all shower parameters including energy and shower maximum. For more inclined events, Tunka-133 provides only the trigger, and a reconstruction of the shower direction.} \label{fig_angularDistribution}
\end{figure}

\section{Event selection}
For analysis of Tunka-Rex measurements we use a modified version of the Offline software framework \cite{RadioOffline2011}, which has been developed by the Pierre Auger Collaboration and now is partially open to other collaborations such as Tunka-Rex. Using the radio modules of this software, we identify all events containing at least 3 radio antennas, which have detected a radio pulse in the time window expected from the air-Cherenkov reconstruction. Since some radio events are contaminated by background pulses, we use only those radio events for analyses, whose reconstructed arrival direction agrees within $5^\circ$ with the direction reconstructed by Tunka-133.

Figure \ref{fig_angularDistribution} shows a skymap of the selected events. Events detected at zenith angles $\theta > 50^\circ$ are marked differently: for these events Tunka-133 cannot provide reliable measurements of the shower energy and the shower maximum, but only of the shower direction. This means that these events are not suitable for the cross-calibration, but potentially can be used for later science analyses.

The skymap reflects two features of the radio technique already known: First, the signal strength increase with the geomagnetic angle, i.e. the angle between the shower axis and the Earth's magnetic field. Thus, the detection threshold is lower for inclined events, since they have large geomagnetic angles. Second, since for inclined showers the shower maximum is more distant. Hence, the radio footprint at ground becomes larger. Thus, sparse radio arrays are more efficient for inclined showers, if the signal has to exceed the detection threshold in at least three antennas. For Tunka-Rex this effect is important, since the antenna spacing is approximately $200\,$m, and for vertical showers the lateral distribution of the radio signal decreases exponentially at distances $\gtrsim 100\,$m.

The combination of both effects makes Tunka-Rex very efficient for inclined showers with a threshold around $10^{17}\,$eV, but less efficient for vertical showers. Still, at the highest energies studied by Tunka, i.e., around $10^{18}\,$eV, the event rates of Tunka-Rex and Tunka-133 are roughly equal. Therefore, as soon as 24/7 measurements become available due to the new Tunka-Grande scintillators, the total statistics of Tunka could be enhances significantly around $10^{18}\,$eV, i.e., exactly in the energy range where statistics is the limiting factor.

\begin{figure}[t]
\centering
\includegraphics[width=0.69\textwidth, angle=-90]{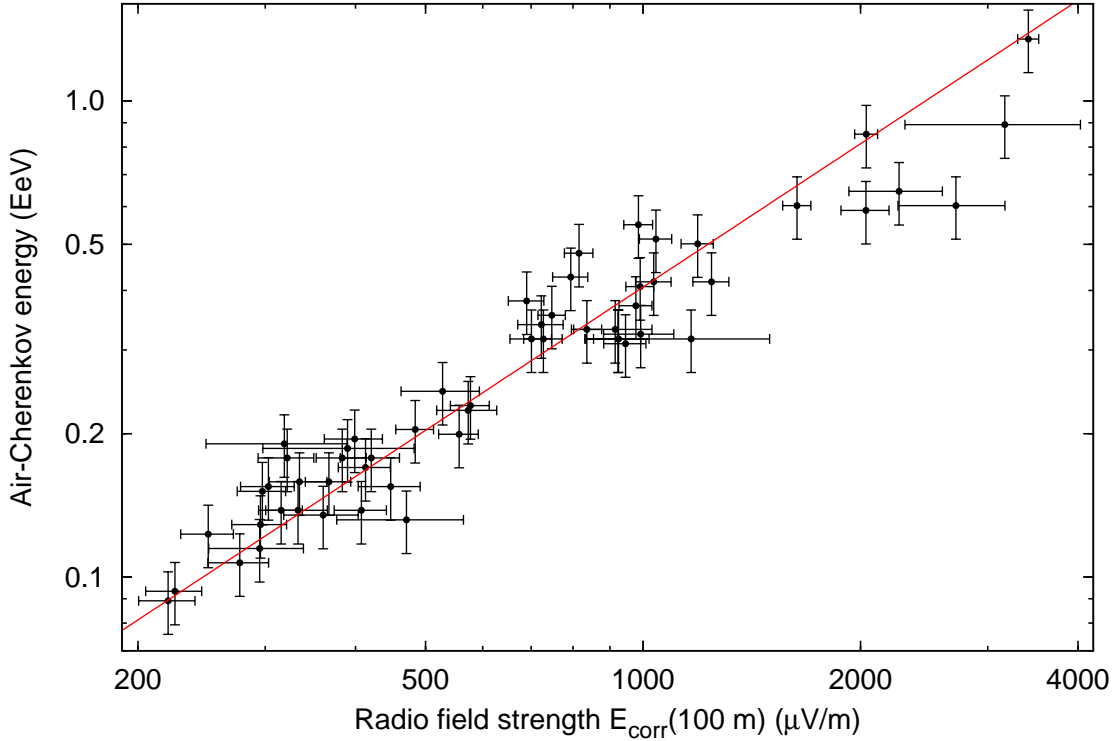}
\caption{Correlation of the energy measured with the air-Cherenkov array and an energy estimator based on the radio amplitude at $100\,$m measured with Tunka-Rex. The line indicates a linear correlation.} \label{fig_energyReconstruction}
\end{figure}

\section{First results}
For determining the precision of Tunka-Rex by the cross-calibration to Tunka-133, we follow a semi-blind approach. Due to the measurement schedule of Tunka-133 breaking in summer, the Tunka-Rex data set is split into two seasons of approximately equal statistics: October 2012 - April 2013, and October 2013 - April 2014. For the first season, the Tunka-133 collaboration has revealed the complete reconstruction to us, including the shower geometry (core and direction), the energy and the shower maximum. For the second season, only the shower geometry is known, which we use to distinguish the radio signal of air shower from background pulses, but the energy and the shower maximum is blind for everybody working on the Tunka-Rex analysis.

The data of the first season is used to optimize and tune the reconstruction of the energy and the shower maximum from the radio measurements. The average deviation between the air-Cherenkov and the radio reconstruction provides an estimate for the precision, and should be minimized. Once the methods are finalized, a prediction will be made for the events of the second season. Afterwards, it will be compared to the air-Cherenkov reconstruction, to check if the average deviation between both reconstructions is still approximately equal to the first season. If not, then likely the methods were 'over-tuned', and the independent check of the second season gives the better estimate for the true precision.

Since the second season is still blind, only results for the first season are presented here. Moreover, we show only the result of the energy reconstruction, since it relies on relatively many events compared to the number of free parameters in the reconstruction method. Thus, the results should be robust, and we expect only little changes for the precision between the 'tuning' and the 'cross-check' seasons. The full result of both seasons, and the results for the shower maximum will be published after the unblinding will have been done.

The energy reconstruction is based on the radio amplitude at a distance of $100\,$m from the shower axis. Results from earlier experiments and simulations \cite{LOPES_XmaxLDF2014, Huege2008, AERAenergy2012} indicate that at this distance the amplitude depends in good approximation only on the geomagnetic angle and the energy, but only little on the shower maximum. Since the geomagnetic angle is known from the reconstruction of the arrival direction, the measured amplitude can be corrected for the strength for the geomagnetic effect. Moreover, we now correct for the known asymmetry of the radio signal due to the interference of the geomagnetic effect and the Askaryan effect \cite{CODALEMAMarinICRC2011,AERA_Polarization2014}. The reconstruction is still based on a simple exponential function for the lateral distribution, since it requires only three antennas with signal. This might explain the outliers for high-energy events, which usually contain more antennas with signal, and would be better fitted by more complex lateral-distribution functions.

Figure \ref{fig_energyReconstruction} shows a comparison for the corrected radio amplitude measured by Tunka-Rex and the energy of the air-Cherenkov reconstruction. The average deviation between the energy reconstructed from the radio and the air-Cherenkov measurements indicates that the precision of the radio measurements is in the same order as the energy precision of the Tunka-133 air-Cherenkov array, which is $15\,\%$. 

\section{Conclusion}
Tunka-Rex demonstrates that a digital antenna array can be a very cost-effective extension for an existing ground array. The energy precision of the radio measurements seems to be competitive to the air-Cherenkov technique, which is slightly better than that of particle detectors. Hence, a radio extension should be of advantage at energies above $10^{17}\,$eV for any science goal relying on energy resolution and high statistics. Moreover, radio measurements provide additional sensitivity to the longitudinal shower development and, thus, a statistical access to the composition of the primary cosmic rays.

Furthermore, radio detection comes with several principal advantages. First, signal attenuation in the atmosphere is negligible, which makes radio interesting for inclined events. Second, radio detection seems to be the only method which provides with full duty-cycle a calorimetric measurement of the electromagnetic component of the air-shower. Since the radio signal is relatively well understood now, there is a chance that even the absolute energy scale becomes competitive with other techniques. At the moment, the absolute scale of the radio amplitude is limited by the calibration of the antennas, which is no principle, but a pure technical limitation, and can easily be improved with reasonable effort.

Consequently, Tunka-Rex seems to be sufficiently mature for the transition from pure technology development to the application of radio measurements for cosmic-ray science. Especially, the additional statistics provided by the inclined Tunka-Rex events, and the future events triggered by Tunka-Grande can be used for physics analyses aiming at the energy range of the assumed transition from galactic to extra-galactic cosmic rays.

\section*{Acknowledgement}
Tunka-Rex has been funded by the German Helmholtz association and the Russian Foundation for Basic Research (grant HRJRG-303). Moreover, this work was supported by the Helmholtz Alliance for Astroparticle Physics (HAP), by the Russian Federation Ministry of Education and Science (agreement 14.B25.31.0010, zadanie 3.889.2014/K), the Russian Foundation for Basic Research (Grants 12-02-91323, 13-02-00214, 13-02-12095, 14002-10002) and the President of the Russian Federation (grant MK-1170.2013.2).

\end{document}